\documentclass[10pt,letterpaper]{article}
\usepackage[top=0.85in,left=1.6in,footskip=0.75in]{geometry}

\usepackage{amsmath,amssymb}

\usepackage{changepage}

\usepackage[utf8x]{inputenc}

\usepackage{textcomp,marvosym}

\usepackage{cite}

\usepackage{nameref,hyperref,soul}

\usepackage[right]{lineno}

\usepackage{microtype}
\DisableLigatures[f]{encoding = *, family = * }

\usepackage[table]{xcolor}

\usepackage{array}

\usepackage{caption,tabularx,booktabs}

\newcolumntype{+}{!{\vrule width 2pt}}

\newlength\savedwidth

\usepackage[aboveskip=1pt,labelfont=bf,font=small,labelsep=period,singlelinecheck=off]{caption}

\makeatletter
\renewcommand{\@biblabel}[1]{\quad#1.}
\makeatother

\usepackage{lastpage,fancyhdr,graphicx}
\usepackage{epstopdf}
\begin{document}
\vspace*{0.2in}

\begin{flushleft}
{\Large
\textbf\newline{\textbf{\textit{\textit{Vizaj}} - An interactive javascript tool for visualizing spatial networks}}
}
%
\bigskip
\\
Thibault Rolland \textsuperscript{1},
Fabrizio De Vico Fallani\textsuperscript{1,*}
\\
\bigskip
\textbf{$^1$} Sorbonne Universite, Paris Brain Institute - ICM, CNRS, Inria, Inserm, AP-HP, Hopital Pitie Salpetriere, F-75013, Paris, France
\\
\bigskip

* Corresponding authors: t.d.rolland@gmail.com; fabrizio.de-vico-fallani@inria.fr

\end{flushleft}
\bigskip

\section*{Abstract}

\begin{small}

In many fields of science and technology we are confronted with complex networks.
Making sense of these networks often require the ability to visualize and explore their intermingled structure consisting of nodes and links.

To facilitate the identification of significant connectivity patterns, many methods have been developed based on the rearrangement of the nodes so as to avoid link criss-cross.
However, real networks are often embedded in a geometrical space and the nodes code for an intrinsic physical feature of the system that one might want to preserve. For these spatial networks, it is therefore crucial to find alternative strategies operating on the links and not on the nodes.

Here, we introduce \textit{\textit{Vizaj}} a javascript web application to visualize spatial networks based on optimized geometrical criteria that reshape the link profiles.
While optimized for 3D networks, \textit{\textit{Vizaj}} can also be used for 2D networks and offers the possibility to interactively customize the visualization via several controlling parameters, including network filtering and the effect of internode distance on the link trajectories.
\textit{\textit{Vizaj}} is further equipped with additional options allowing to improve the final aesthetics, such as the color/size of both nodes and links, zooming/rotating/translating, and superimposing external objects.

\textit{\textit{Vizaj}} is an open-source software which can be freely downloaded and updated via a github repository. Here, we provide a detailed description of its main features and algorithms together with a guide on how to use it.
Finally, we validate its potential on several synthetic and real spatial networks from infrastructural to biological systems.

We hope that \textit{\textit{Vizaj}} will help scientists and practitioners to make sense of complex networks and provide aesthetic while informative visualizations.

\end{small}

\hspace{1cm}

\hspace{1cm}

\textbf{Keywords} - Complex systems, Physical networks, Dataviz, Software, Art


\newpage

\section*{Introduction}

Networks are abstract representation of interconnected systems, consisting of nodes (ie, the units) and links (ie the connections).
Many real systems can be modeled and analyzed through networks from social and technological networks to biological and medicine networks \cite{newman_structure_2003,boccaletti_complex_2006}.

In the last decades, the development of theoretical means and availability of more and more accurate data has literally boosted the broad interest in complex networks \cite{costa_analyzing_2011}.
Understanding complex networks requires the use mathematical languages, such as graph theory, data mining and statistics, to analyze and model the underlying structural properties and relate them to the dynamics of the system \cite{latora_complex_2017,cimini_statistical_2019}.
Nonetheless, the visualization of these networks is perhaps the first entry point to gain intuition on the main connection features of the system under investigation. As such, network visualization is becoming an important related field of network science as well as of modern art, providing intelligible and aesthetic patterns, too \cite{barabasi_network_2016}.

Stemming from the \textit{dataviz} field, network visualization aims to optimize the information that we can obtain by visually inspecting the graph.
While there are no strict criteria for aesthetics of a drawing in both 2D or 3D, it is generally agreed that such a drawing has minimal edge crossing, with nodes evenly distributed in the space, connected nodes close to each other, and symmetry that may exist in the graph preserved.
To this end, many algorithms have been developed based on different criteria such as the spring-electrical models, the stress and strain models, a well as high-dimensional embedding and Hall's algorithms \cite{Herman_graph_2000}. 
Those methods can be found in many existing software such as Gephi, Cytoscape or Pajek just to cite a few \footnote{\url{https://gephi.org}, \url{https://cytoscape.org}, \url{https://mrvar.fdv.uni-lj.si/pajek/}}.
The main working strength of all these methods is the possibility to freely rearrange the node positions so as to optimize some quality function.

However, many real networks are spatially embedded and their fixed node location actually conveys important constraints to the resulting topological structure \cite{barthelemy_spatial_2011}. This is intuitively related to the cost needed to establish long distant connections in the system. For example, it has demonstrated how brain systems optimize the trade-off between the network efficiency and the metabolic energy needed to build long axonal connections \cite{bullmore_economy_2012}. 
Hence, in spatial networks the position of the nodes is a feature that we might not want to visually alter. Although a similar reasoning could be made for links too, in many situations they do not actually represent physical quantities but they are rather inferred from the data via some statistical procedures. This is the case for example for many social, biological and finance systems, where networks have to be reconstructed from the local information of the nodes \cite{brugere_network_2018}.

Shaping the links so as to avoid many intersections while keeping fixed the nodes is a very hard problem and some recent attempts have been made in physical networks inspired by self-avoiding polymer chains and manifold dynamics \cite{dehmamy_structural_2018}. 
While the results show the effectiveness in avoiding crossing conditions the underlying algorithms are computationally heavy and might give many tortuous paths thus making hard to make sense of the network.
In an effort to find a lighter and flexible solution we introduce simple geometric criteria that alter the shape of the links based on the relative Euclidean distance between nodes. 
We embed these rules in a new web-browser, general-purpose and interactive network visualization tool that we named \textit{Vizaj}.

\textit{Vizaj} is designed to work with spatial networks in both 2D and 3D. It has many parameters allowing the user to customize the visualization and it does not require any installation. \textit{Vizaj} only requires as input an adjacency matrix corresponding to an (un)weighted graph and the positions of the nodes. All the output visualizations can be then saved and exported as high-quality pictures and reusable 3D models.
In the following, we show the working principles of \textit{Vizaj}, explain its controlling parameters, and validate its potential on synthetic and real networks from social to biological systems.

\textit{Vizaj} can be directly used at \url{https://bci-net.github.io/vizaj/}. Being an open source project, \textit{Vizaj}'s code and sample data can be freely accessed at the github page \url{https://github.com/BCI-NET/vizaj}.

\section*{Design and implementation}

In this section, we first describe the geometrical criteria introduced to shape the link paths in a spatial network and then provide an overview of the software implementation as well as a description of the available controlling parameters.

\subsection*{Geometric criteria for shaping network links}

\textit{Vizaj} operates within a 3D environment where links lie on planes defined by the spatial position of two connected nodes $S_i$ and $S_j$ and by a virtual reference point $C$, which is initially set to the barycenter of all the node positions, but can be next arbitrarily moved along the altitude axis. 

Accordingly, we define a virtual reference line identified by $C$ and by the point $M$ falling in the middle of the geodesic between $S_i$ and $S_j$. 
The links are made up of two quadratic Bezier curves starting respectively from the two nodes and reaching a summit point $U$ on the reference line . 
The vertical position of $U$ is always contralateral to $C$ so as to generate concave-like link shapes. Each Bezier curve has two boundary points $B_1$ abd $B_2$ whose positions determine the link shape (\textbf{Fig. 1}a).

The exact position of the Bezier boundary points are indirectly defined following geometric rules that take into account the relative Euclidean spatial distance $d_{ij}$ between the nodes and the angle $\alpha$ of the curve entering into each node.
Namely, there are four parameters that the user can control (\textbf{Fig. 1}b):

\begin{itemize}

\item{The height \(h\), corresponding to the distance between the \(U\) and \(M\). This is proportional to the internode distance, ie $h=a_1d_{ij}$}

\item{The distance $d_U$ between the top point $U$ and $B_2$ (or $B_2’$). This is proportional to the internode distance divided by 2, ie $d_U=a_2\frac{d_{ij}}{2}$}.

\item{The node angle $\alpha$, which sets the orientation of the curves entering into the nodes. This is only proportional to the flat angle, ie $\alpha=a_3\pi$}.

\item{The distance \(d_S\) of the node handle between \(S_i\) and \(B_1\) (or \(S_j\) and \(B_1’\)). This is also proportional to the internode distance divided by 2, ie $d_S=a_4\frac{d_{ij}}{2}$}.

\end{itemize}

By manipulating the $a_1, a_2, a_3, a_4$ coefficients, different link shapes can be generated resulting in a multitude of possible aesthetic configurations (\textbf{Fig. 1}b).

Note that the radius of the links automatically decreases with the internode distance $d_{ij}$ so that the highest links are also the thinner ones. This improves the visibility of the spatially shortest connections otherwise hidden by the longest ones.

\subsection*{Graphical user interface and main features }

\textit{Vizaj} is intended to be lightweight tool that can be run without any installation. 
To this end, we coded \textit{Vizaj} in javascript using the Three.js framework (\url{https://three.js}). 
Three.js is a GPU-accelerated 3D processing javascript language compiled with WebGL, a low-level graphics API designed specifically for the web. 
Notably, \textit{Vizaj} is hosted online, which makes it usable on any machine through internet, and it runs standalone without relying on proprietary browser plugins.

A dedicated graphical user interface (GUI) facilitates the interactive manipulation and creation of the visualization by the end user (\textbf{Fig. 2}).
The whole process consists of three steps: \textit{i)} the data import, \textit{ii)} the network visualization, and \textit{iii)} the export of the final outcome.
A detailed guide is also provided in the Github page \url{github.com/BCI-NET/vizaj}

\subsubsection*{Data import}

The network data needs to be accessible from the web browser. 
Two types of information must be provided, i.e. the 3D spatial position of the nodes and the adjacency matrix corresponding to the  network. The typical input data consists of connected weighted networks. \textit{Vizaj} currently does not work with directed networks. Node labels can be also provided, albeit they are optional.
\textit{Vizaj} offers the possibility to upload those information through standard .csv files or encapsulated .json files.
Note that the axis orientation is xzy (and not xyz). Thus, the input data should respect this order to correctly visualize the network in its actual space.

\subsubsection*{Network visualization}

Once the input data are uploaded, the network is immediately shown in a 3D environment.  Nodes are made up of spheres and each link is generated as a curve between its endpoints. 
By default, the nodes and their links are dynamically highlighted when hit by the cursor. Labels also pop out if given as input.
To ensure an efficient visualization, networks are initially filtered so as to keep a small percentage the strongest links via the ECO criterion \cite{de_vico_fallani_topological_2017}. 
The user can then tune the parameters in the GUI to obtain alternative network visualizations. 
Several levels of customization are provided including the network density, the geometry, radius and color of the links/nodes and of the background.
In order to improve the visual perspective, 3D items can be added to the scene, such as a brain (default), a sphere or a cube. They can be moved, rotated and scaled using a mesh helper in the GUI. 
In addition, information about the number of links for each node can be visually shown as cylinder along the $UM$ axis whose height is proportional to the node degree.

\subsubsection*{Graphical export}
 
Once correctly tuned, the final network visualization can be saved either as a 3D object or as a picture.
The 3D object is exported with the standard file format for three-dimensional scenes and models, ie GL transmission format (.gltf).
This file can then be used in 3D modeling tools such as Blender, Unity, or other Three.js based scripts \footnote{\url{https://blender.org}, \url{https://unity.com}}.
The picture format is .tiff with a $96$ DPI resolution and horizontal and vertical dimension of $6000$ and $3500$ pixels, respectively.
Note that node labels are not saved in the picture, but only visualized in the website when hit by the cursor. 

\section*{Results}

Here we present a collection of results obtained with different spatial networks. 
To illustrate the potential of \textit{Vizaj} we both analyzed synthetic and real networks, and discuss the effect of the chosen parameter values with respect to the nature of the data.
We finally present a brief overview of \textit{Vizaj} in terms of its computational complexity and performance.

\subsection*{A configurable interactive visualization}

In general, network links are typically given a size which is proportional with their weight, i.e. the intensity of the connection. With spatial networks, this choice might actually hamper the visualization of the weakest links. This is further complicated by the possible existence of long connections which typically bring confusion and give messy results.

To overcome this limitation, \textit{Vizaj} adopts a two-fold strategy. On one hand, the size of the links is only proportional to the inverse of the distance they cover. 
On the other hand, the height of the links is directly proportional to covered distance (\textbf{Fig. 3}). Note that the color of the links is instead coding for the weight of the connection.
These features are particularly suitable for interconnected systems exhibiting a large variability in the internode spatial distance, such as city transport networks (\textbf{Fig. 4}).

Another popular graphical convention is to show the importance of the nodes by setting their size proportional to the node degree, ie the number of links connected to it. 
However, in the case of spatial networks, this might lead to overlapping nodes given that there is a limited available space they can occupy. 
\textit{Vizaj} circumvents this problem by representing node degrees as vertical lines perpendicular to the spatial surface (\textbf{Fig. 5}).
This graphical solution is particularly useful when dealing with nodes lying on simple surfaces (eg, flat or convex) such as the worldwide airline route network (\textbf{Fig. 6}).

Finally, it is often desirable to represent the network over an actual physical object to give a more realistic and clearer idea of its spatial organization.
\textit{Vizaj} offers the possibility to visualize the network together with a 3D support that facilitates the threedimensional visual perception of the interconnected system (\textbf{Fig. 7}).
This feature is, for example, useful when representing biological systems whose network is inferred from sensors corresponding to precise sites of an actual physical organ such as the brain (\textbf{Fig. 8}).

\subsection*{Performance}
The overall performance depends on the number of nodes and links in the network, as well as on the type of visualization of the links. Links can be drawn as \textit{lines} made of adjacent points of $1$ pixel size, or as \textit{volumes} made of adjacent small cylinders (\textbf{Fig. 9}). 
The line option is always more efficient than the volume option in terms of G/CPU resources, although the the latter can give more flexible and pleasant results.

We tested the performance with a $2$ GHz $4$ core CPU, $16$GB of RAM, and a $1536$ Mo GPU. 
With the \textit{line} option (\textbf{Fig. 9}a), we could draw networks of $250$ nodes and $31125$ links within few seconds. When using the \textit{volume} option (\textbf{Fig. 9}b), the time needed to visualize the same network was on the order of dozens of seconds.
As the number of nodes becomes higher than $250$ the performance starts to deteriorate progressively leading to significantly longer delays. This is particularly problematic for the smooth usage of the GUI, as many manipulations and customization require the redrawing of all the links.

While some improvements can be done in the future by optimizing our code, we notice that there are intrinsic limitations in the three.js code that one cannot directly modify.
As our ultimate goal was to generate high-quality static visualizations, \textit{Vizaj} can still be considered a useful tool even in the case of very large spatial networks.

\section*{Conclusions}

\textit{Vizaj} was designed to fill in the gap of spatial network visualization in 3D.
By preserving the actual location of the nodes in the network, while changing the way links can be drawn, the user can have a more intuitive  interpretation of the organizational features of the system under investigation.
As such, we hope that \textit{Vizaj} can be used extensively by a broad community of scientists and practitioners, so as to facilitate the communication of their results as well as provide high-quality artistic representations. 

Note that \textit{Vizaj} can also be used to visualize non-spatial networks provided that a node position is given as input, eg precomuted from standard layout algorithms. Currently, \textit{Vizaj} is not optimized to automatically process and visualize huge dataset samples. However, it can be easily embedded in other online platforms designed to perform big data processing, such as Brainlife for the fast and reproducible analysis of neuroimaging data\footnote{\url{https://brainlife.io}}.

\textit{Vizaj} was designed as an open source project and its usage can be acknowledged by citing this article. \textit{Vizaj} is continuously updated with new features and regularly maintained. Suggestions of improvements, as well as further developments, can be addressed to the corresponding authors of this article.

\section*{Acknowledgments}
The research leading to these results has received funding from the ``Agence Nationale de la Recherche''
through contract number ANR-15-NEUC-0006-02.
The content is solely the responsibility of the authors and does not necessarily represent the official views of any of the funding agencies.

\nolinenumbers

\clearpage
\bibliographystyle{plos2015}
\nolinenumbers
\begin{small}
\bibliography{Fab}
\end{small}

\newpage
\section*{Figures}

\begin{figure}[ht!]
    \centering
    \includegraphics[width=\textwidth]{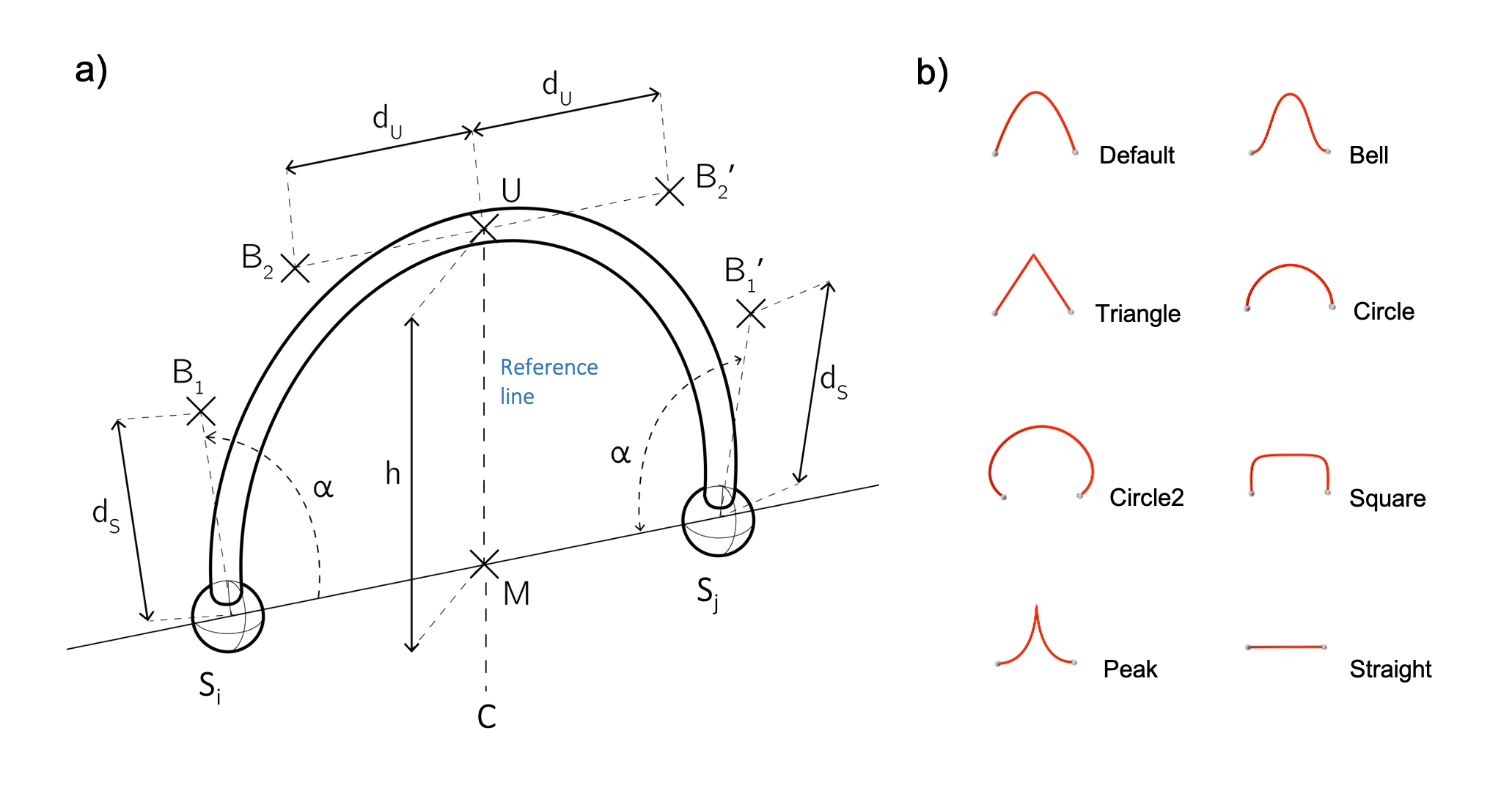}
    \caption{\textbf{Geometric principles for shaping links}. a) Schematic illustration of the parameters controlling the shape of the link between two nodes $S_i$ and $S_j$. The reference point $M$ lies on the geodesic distance $d_ij$ between the nodes. Bezier boundary points are referred as to $B$ and can be moved by modulating four quantities, ie $h$, $d_U$, $\alpha$ and $d_S$. Note that the angles \(\widehat{B_2UM}\) and \(\widehat{B2'UM}\) are always square and not customizable for the sake of simplicity.
    b) Typical configurations that can be obtained with different combinations of the controlling parameters $a_1$, $a_2$, $a_3$, $a_4$ in the GUI. Default (0.75, 0.5, 0.38, 0), Bell (0.75, 0.5, 0, 0.5), Triangle (0.75, 0, 0, 0), Circle (0.5, 0.5, 0.5, 0.5), Circle2 (0.9, 1, 0.8, 1), Square (0.5, 1, 0.5, 1), Peak (0.75, 0, 0, 1), Straight (0, 0, 0, 0).}
    \label{fig:1}
\end{figure}

\newpage
\begin{figure}[ht!]
    \centering
    \includegraphics[width=\textwidth]{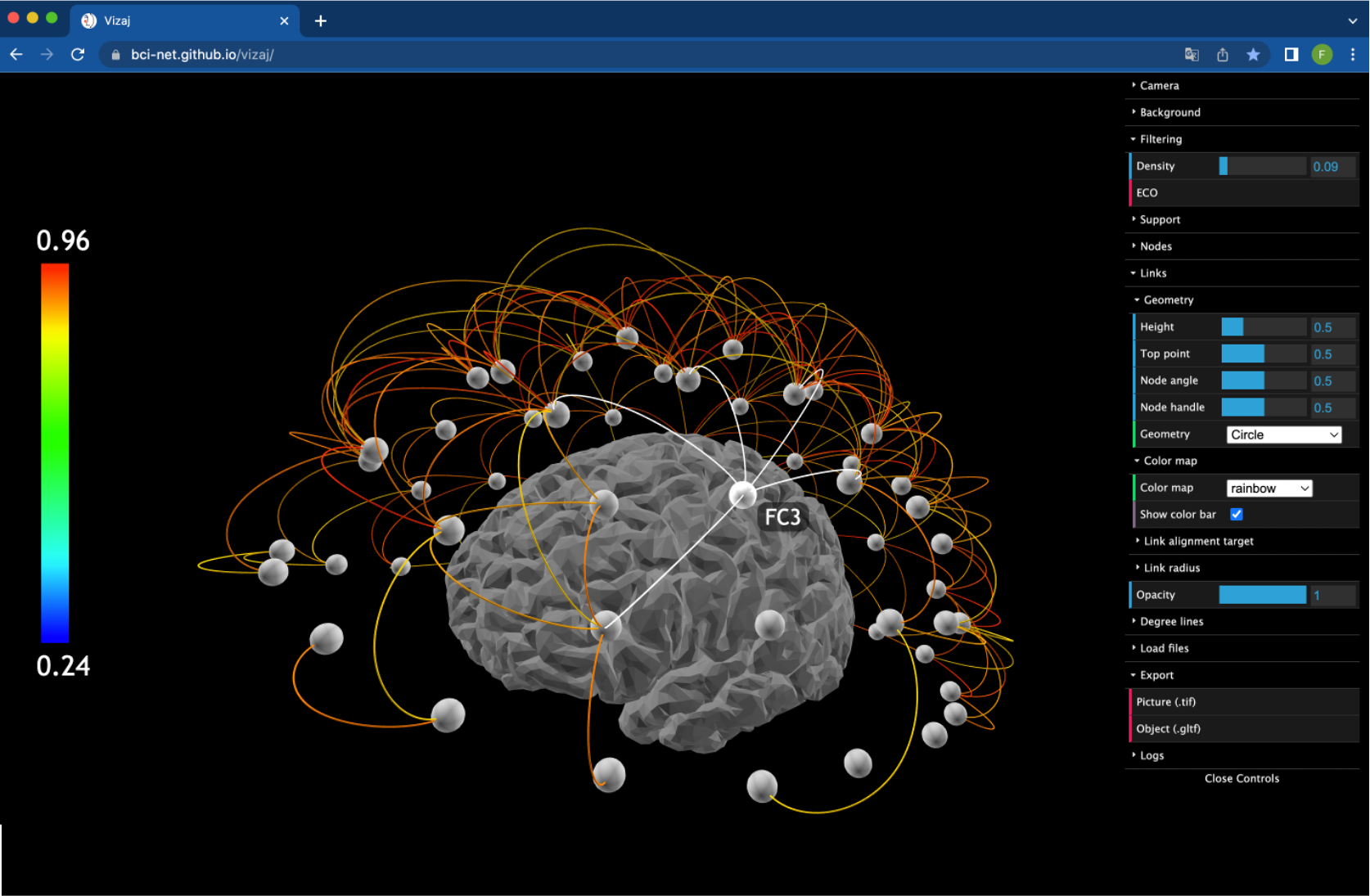}
    \caption{\textbf{Screenshot of the \textit{Vizaj}'s graphical user interface (GUI).} Spatial networks are built in a 3D scene. The network can be zoomed and rotated via the cursor input. When the cursor hit a node, its connections and label are dynamically highlighted. On the right, all the controlling parameters are listed and accessible to the user. }
    \label{fig:2}
\end{figure}

\newpage
\begin{figure}[ht!]
    \centering
    \includegraphics[width=6cm]{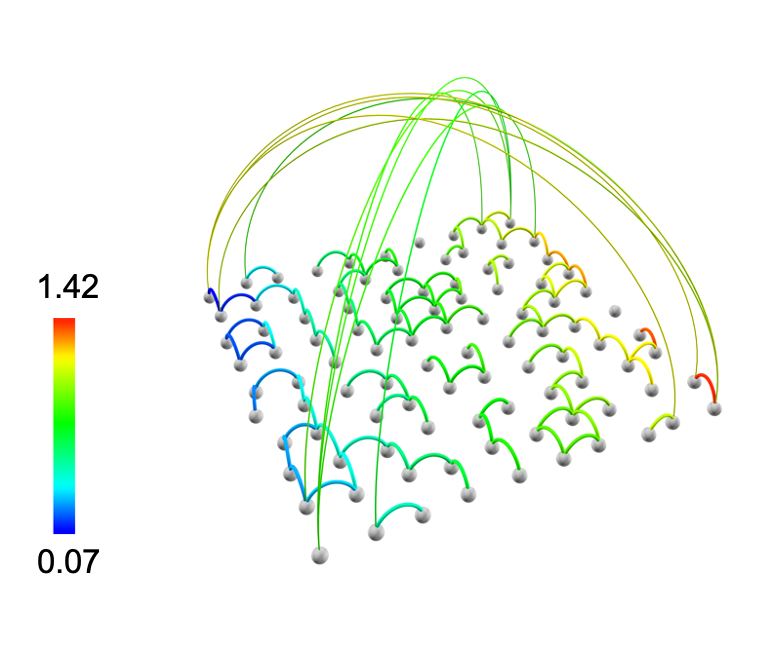}
    \caption{\textbf{Height, radius and color of the links}. The height of the links is proportional to the internode distance $d_{ij}$. The radius is instead inversely proportional to $d_{ij}$. The color of the links code for the weight of the connections. This synthetic network has been visualized with \textit{Vizaj} using a Circle shape for the links.}
    \label{fig:3}
\end{figure}

\begin{figure}[ht!]
    \centering
    \includegraphics[width=\textwidth]{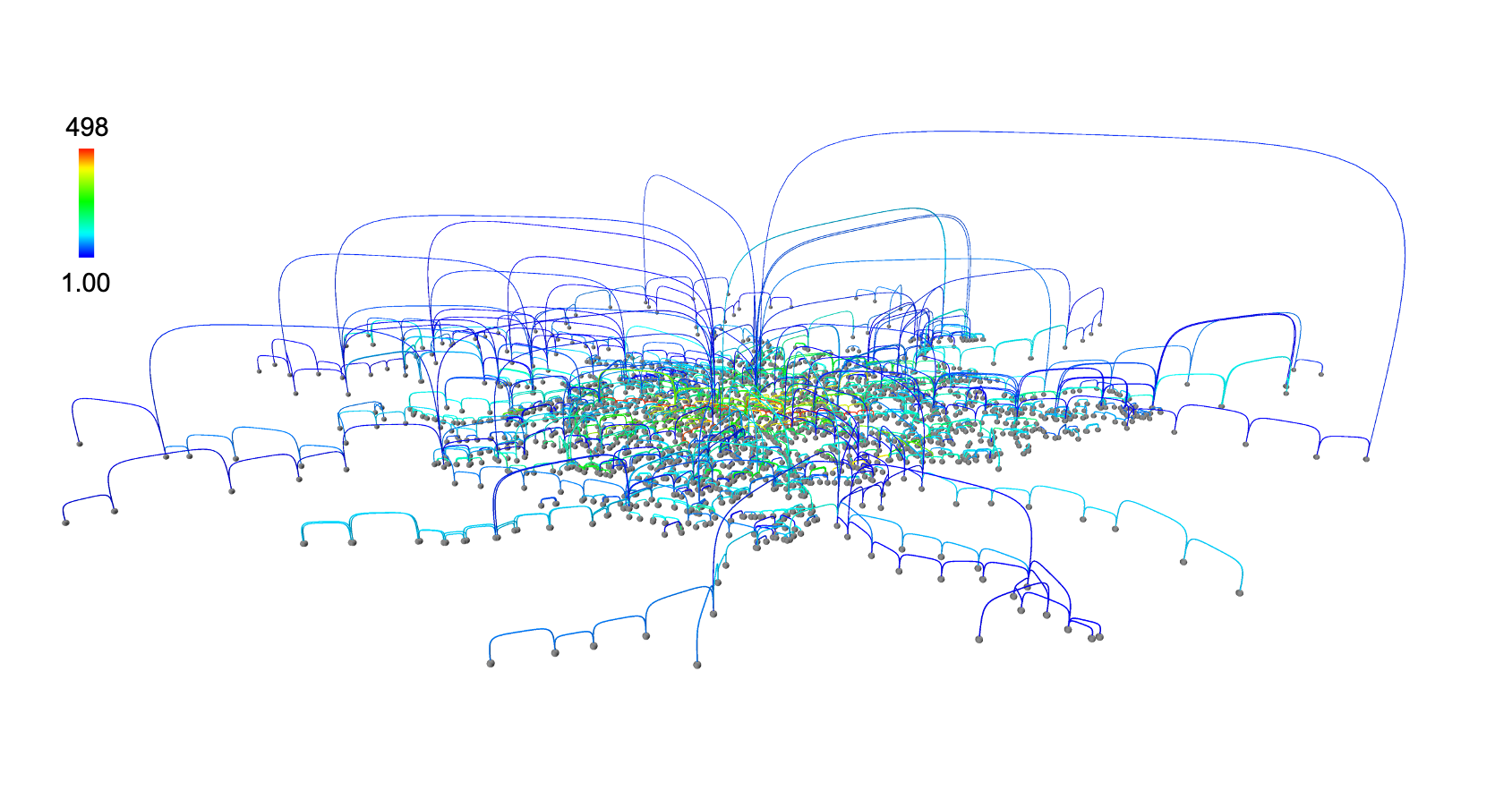}
    \caption{\textbf{Paris public transports.} Nodes are stops and links is travel connections. Mixed modes of travel are included (tram, rail, metro and bus). The color of the links indicates the number of vehicles that used that connection on a Monday. Only the connections covering a physical distance above $500$ meters are displayed for the sake of simplicity. Data available from \cite{kujala_collection_2018}. The network visualization has been generated with \textit{Vizaj} using a Square shape for the links.}
    \label{fig:4}
\end{figure}

\newpage
\begin{figure}[ht!]
    \centering
    \includegraphics[width=6cm]{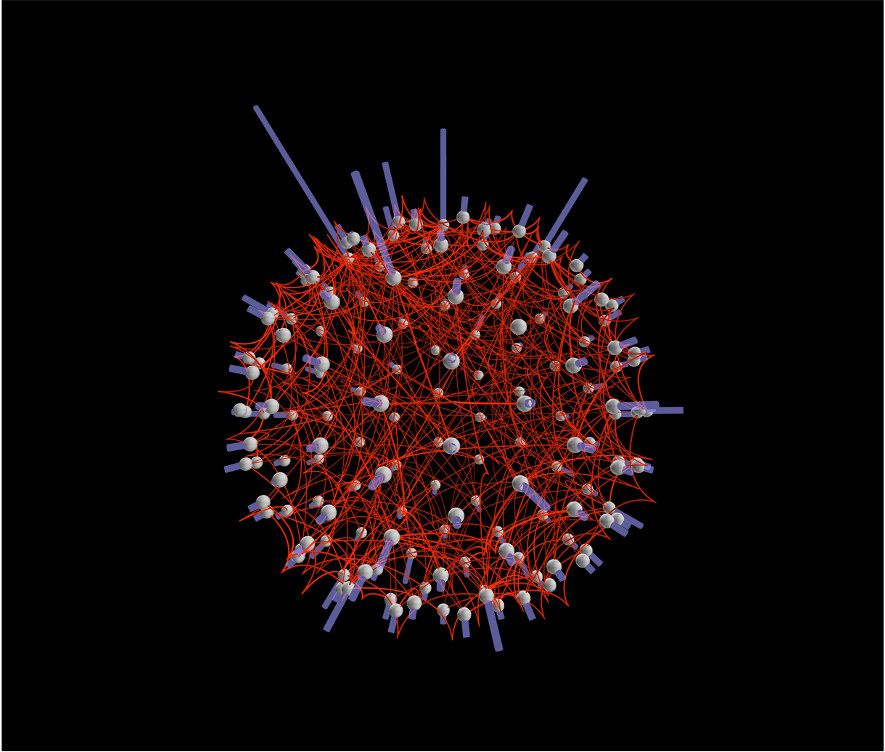}
    \caption{\textbf{Representation of the node degrees}. In this synthetic network, nodes are homogeneously distributed on a sphere. Links are established via a Barabasi-Albert model. Node degrees are illustrated as vertical bars perpendicular to the surface of the nodes. The network visualization has been generated with \textit{Vizaj} using a Peak shape for the links.}
    \label{fig:5}
\end{figure}

\begin{figure}[ht!]
    \centering
    \includegraphics[width=\textwidth]{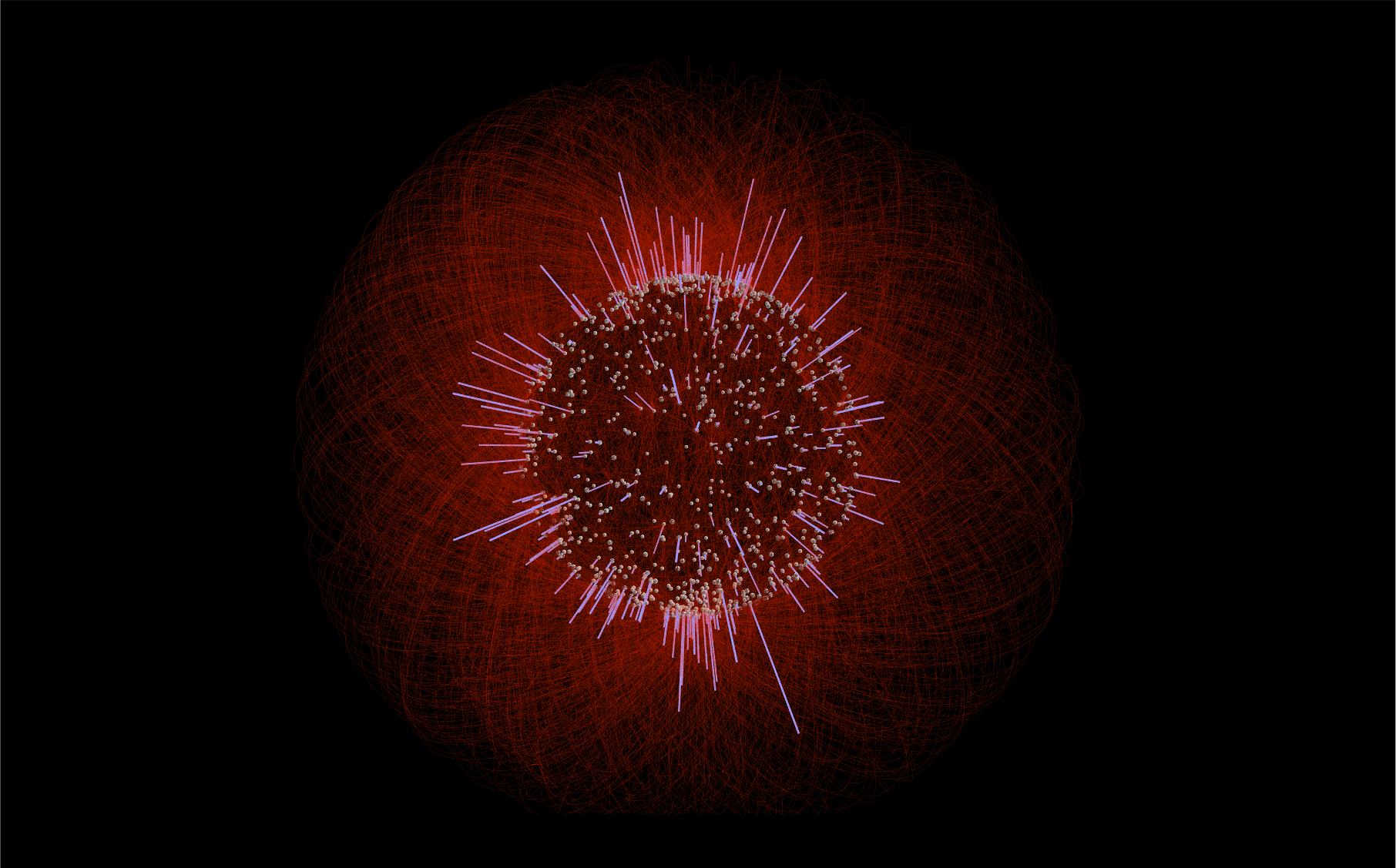}
    
    \caption{\textbf{Worldwide airline routes}. Nodes correspond to airports, links correspond to flight routes. Only the $1000$ most conencted nodes are shown for the sake of simplicity. Data available from \url{openflights.org/data.html}. The network has been visualized with \textit{Vizaj} using a Circle shape for the links.}
    \label{fig:6}
\end{figure}

\newpage
\begin{figure}[ht!]
    \centering
    \includegraphics[width=6cm]{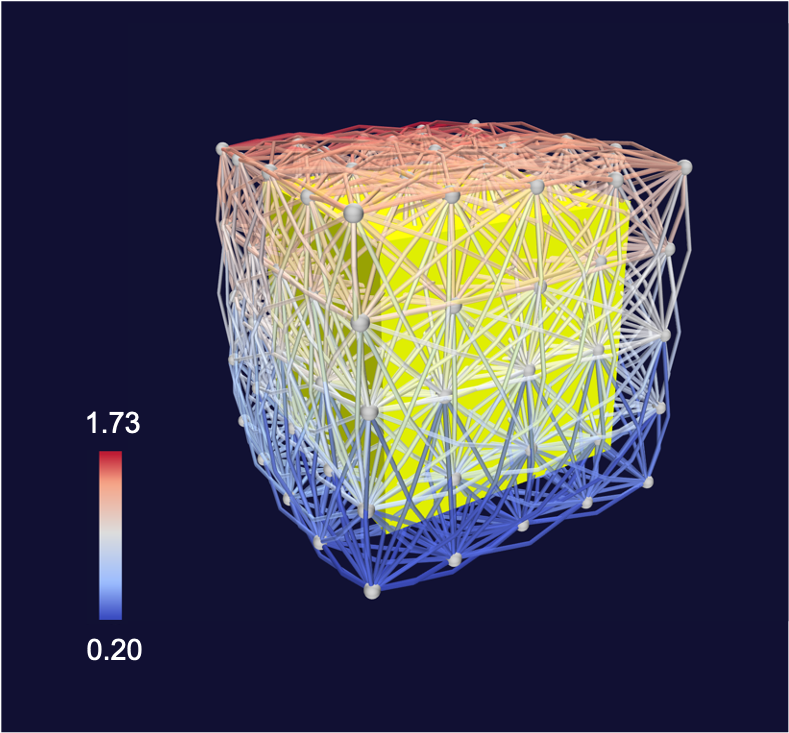}
    \caption{\textbf{Visualization of 3D supports}. A yellow cube is shown within a 3D synthetic lattice. Links are mostly established between neighbor nodes. The color of the links indicates the elevation of the connection. The network has been visualized with \textit{Vizaj} using a Triangle shape for the links.}
    \label{fig:7}
\end{figure}

\begin{figure}[ht!]
    \centering
    \includegraphics[width=\textwidth]{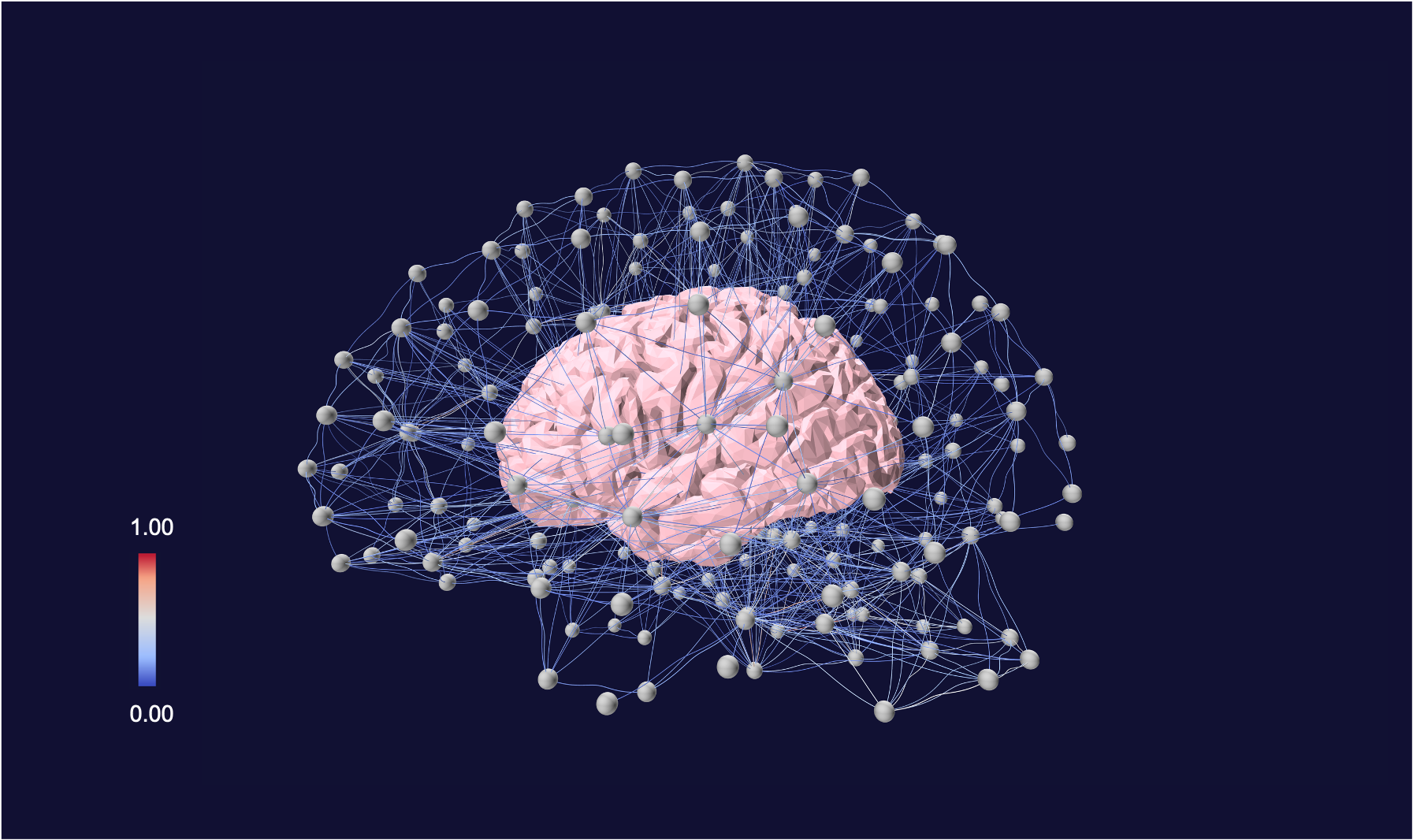}
    \caption{\textbf{Structural brain connectivity}. Nodes represent brain regions of interest, links represent anatomical connections between nodes. In particular, the color of the links indicates the probability to find anatomical fiber bundles between different brain areas. Data available from \cite{brown_ucla_2012}. The network has been visualized with \textit{Vizaj} using a Bell shape for the links.}
    \label{fig:8}
\end{figure}

\newpage
\begin{figure}[ht!]
    \centering
    \includegraphics[width=\textwidth]{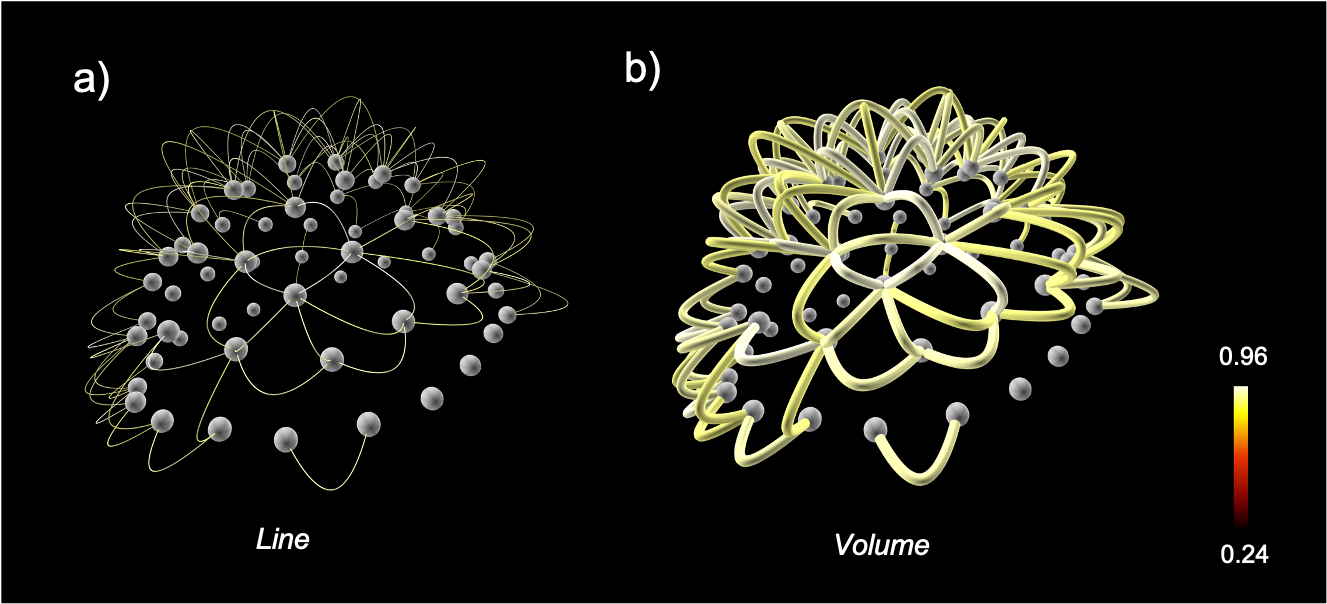}
    \caption{\textbf{Graphical representations of the links} a) Line option: links are made of single pixels. b) Volume option: links are made of multiple adjacent cylinders along the path between the two connected nodes. This synthetic network has been visualized with \textit{Vizaj} using a Default shape for the links.}
    \label{fig:9}
\end{figure}

\end{document}